\documentclass{PoS}
\usepackage{ctable}

\title{Neutrino Sources and Properties}

\ShortTitle{Neutrino Sources and Properties}

\author{\speaker{Francesco Vissani}\\
        INFN, Gran Sasso Science Institute \& Laboratori Nazionali del Gran Sasso\\
        E-mail: \email{vissani@lngs.infn.it}}

\abstract{
In this lecture, prepared for PhD students,
basic considerations on neutrino interactions, 
properties and sites of production are overviewed.  
The detailed content is as follows: 
{\em Sect.~1,
 Weak interactions and neutrinos:} 
{Fermi coupling; definition of neutrinos; global numbers.}
{\em Sect.~2, 
A list of neutrino sources:}
{Explanatory note and examples (solar pp- and 
supernova-neutrinos).}
{\em Sect.~3, Neutrinos oscillations:}
{Basic formalism (Pontecorvo);
matter effect (Mikheev, Smirnov, Wolfenstein);
status of neutrino masses and mixings.}
{\em Sect.~4, Modifying the standard model to include neutrinos masses:} {The fermions of the standard model; 
one additional operator in the standard model (Weinberg); implications. One summary table and several  exercises offer the students occasions to check, consolidate and extend their understanding; the brief reference list includes historical and review papers and some entry points to active research in neutrino physics.}}

\FullConference{Gran Sasso Summer Institute 2014 Hands-On Experimental Underground Physics at LNGS - GSSI14,\\
		22 September - 03 October 2014\\
		INFN - Laboratori Nazionali del Gran Sasso, Assergi, Italy }

\begin{document}

\section{Weak interactions and neutrinos \label{s1}}
We assume that the reader has some acquaintance with the standard model of electroweak interactions. We use this assumption to 
introduce the topics of interest 
as effectively as possible. 
\subsection{Fermi coupling}
The hamiltonian that causes the weak charged-currents 
transitions is, 
\begin{equation}
H_{cc}=\frac{G_F}{\sqrt{2}}
\int d^3x \ (J_{\mbox{\tiny cc}}^\mu)^\dagger J_{\mbox{\tiny cc},\mu}
\mbox{ where we sum over }\mu=0,1,2,3
\end{equation}
The factor $\sqrt{2}$ is purely conventional. The amplitude of any weak transition at low energy is proportional to the Fermi coupling $G_F$, thus any decay width or cross section is proportional to its square. (We discuss later some 
physical quantities that are linear in $G_F$.) 
The  numerical value is,
\begin{equation}
G_F^2= 5.297\times 10^{-44}\  \frac{\mbox{cm}^2}{\mbox{MeV}^2}
\end{equation}
where we have used $\hbar c\sim 200$ MeV fm, with 1 fm=$10^{-13}$ cm. 
From the point of view of the standard model, the above interaction derives from the   (tree-level) exchange of a $W$ boson in the low energy limit: Thus, we get \begin{equation}
\frac{G_F}{\sqrt{2}}=\frac{g^2}{8 M_W^2}
\label{fdef}\end{equation} 
where the value of the $W$ mass is $M_W\sim 80$ GeV and its coupling to the fermions is $g\sim 0.65$.

\subsection{Definition of neutrinos}
The weak charged current $J_{\mbox{\tiny cc}}^\mu$  decreases the 
electric charge of the fermionic state by one unit. It  
contains two parts, one leptonic and one hadronic (or in fact `quarkonic'). The first one is,
\begin{equation}
J_{\mbox{\tiny cc, lept}}^\mu/2=\bar{e} \gamma^\mu P_L \nu_e + 
\bar{\mu} \gamma^\mu P_L \nu_\mu + 
\bar{\tau} \gamma^\mu P_L \nu_\tau 
\end{equation}
where $e,\mu,\tau$ and $\nu_e,\nu_\mu,\nu_\tau$ are relativistic quantum fields,  and where the chiral projector 
 $P_L=(1-\gamma_5)/2$ selects the states with left chirality. 
 This current 
provides us with the {\em definition} of neutrino `species' (or `type', or `flavor'): the electronic neutrino field is the one associated to electronic field, and similarly for the other ones. Thus, by definition, the neutrino emitted in the pion decay
$\pi^+\to \mu^+ + \nu_\mu$ is a muonic neutrino, whereas the one emitted in the beta decay of the neutron $n\to p+ e^-+ \bar\nu_e$ is an electronic antineutrino.
In the standard model neutrinos are massless, that agrees with the observation that 
neutrinos have negative helicity and antineutrinos positive helicity.

\subsection{Global numbers}

Experience shows that,  in any known reaction, 
the number of leptons of any type does not change.
This leads to the conclusion that the electronic lepton number $N_e$,    
the muonic lepton number $N_\mu$ etc, are conserved, just as the total number of leptons,  
\begin{equation}N_L=N_e+N_\mu+N_\tau\end{equation} 
or the total number of baryons $N_B$. This observational fact is  neatly accounted for in the standard model, since  there are  
global symmetries associated to 
conserved currents, e.g.\ 
$d J_e^\mu/dx^\mu=0$, 
where the time-component of the 
electronic current is 
$J_e^0=e^\dagger e+\nu_e^\dagger P_L \nu_e$ and 
$N_e=\int  J_e^0(t,\vec{x})\, d^3 \! x$.

\begin{figure}[t]
\begin{minipage}{10cm}
 \caption{\small\em The baryon and the lepton numbers are conserved in the classical theory: their currents obey $d  J_\mu/dx_\mu=0$.
Instead, in quantum field theory the divergence is non-zero when we consider loop diagrams 
with $W$ bosons in external states (`anomaly'). This leads to  transitions that change $B$ and $L$, when $W$ are in thermal equilibrium.   \label{tri}}
 \end{minipage}
 \hfil
   \begin{minipage}{5cm}
   \includegraphics[width=.9\linewidth]{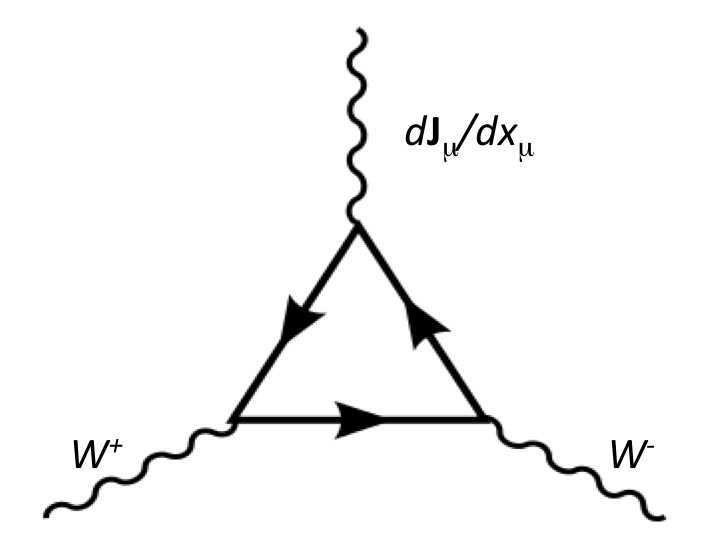} 
   \end{minipage}
\end{figure}

For completeness, note that the last statement is  true only neglecting quantum fluctuations. Indeed, strictly speaking 
$N_L$ and $N_B$ (the baryon number) are not respected in the 
standard model, 
since the divergence of the leptonic and baryonic currents are not zero. E.g.\ the diagram of Fig.~\ref{tri} yields
the contribution,  
\begin{equation}\frac{d J_L^\mu}{dx^\mu}=
\frac{d J_B^\mu}{d x^\mu}=\frac{3 g^2}{32 \pi^2} \mbox{Tr}[F_{\mu\nu} \widetilde{F}^{\mu\nu}]\end{equation}
Thus the divergence is non-zero and the global numbers are violated. But these effects  
become conspicuous only when $W$-bosons are copiously produced, e.g.\ at high temperatures that occur in the early Universe.
In the following, we will not 
develop these remarks further
and 
focus on phenomena that are directly 
observable in terrestrial laboratories.

{\small
\vskip2mm
\leftline{\bf Exercises of Sect.~\ref{s1}}
\vskip1mm
\sf
\noindent 1) Check the dimensions of all formulae in these notes, by using the rules prescribed by so-called system of natural units (i.e.\ the system used in particle physics where $\hbar=c=1$) namely:
\begin{quote} 
length=time=1/mass=1/momentum=1/energy.
\end{quote}
\noindent 2)
Write down the (leptonic and hadronic) currents that appear in 
electromagnetic, charged current and neutral current interactions.
Discuss their similarities and differences.\\
\noindent 3) Find/guess which are combinations of global numbers that 
are conserved in the standard model even accounting for quantum fluctuations. Can they be promoted to new gauge symmetries?
}

\section{A list of neutrino sources \label{s2}}
In table~\ref{tab1}, that appears at the end of these note, 
several neutrino sources are considered. For each of them, we list various features, including the most important one: the number of observed (or potentially observable) events. 
This is given by
\begin{equation}
\mbox{events}=\mathcal{N}\times T\times \Phi \times \sigma
\end{equation}
where the 4 terms in the r.h.s.\ are the number of targets $\mathcal{N}$, the time of data taking $T$, the flux 
$\Phi$ and the cross section of the relevant interaction $\sigma$.
Few experiments are mentioned in table~\ref{tab1}, but making reference only to the number of relevant targets and to the time of data taking.

\subsection{Explanatory note}

This table is useful for a first orientation. We fixed the value of the total cross section at some relevant energy, and then checked that with a suitable average value of the flux, this gives the correct (measured or expected) number of events. 

Often, the most relevant quantity is the number of events. 
Note that in the scientific literature the time $T$ is 
 sometimes included in the 
{\em fluence} $F$, i.e.\ the time integrated flux
\begin{equation}
F=\int  \Phi(t)\, dt=\Phi\times T
\end{equation}
where $\Phi$ in the r.h.s.\ is the time-averaged flux.
Alternatively, the time is included  
in the {\em exposure}, namely the product
$\mathcal{N}\times T$.
Another combination that is commonly used is the {\em effective area} given by $\mathcal{N} \times \sigma$.
We do not take into account explicitly any efficiency factor, that can be attached to the effective cross section or to the exposure or to the effective area. Note 
finally that in the table and in our simplified estimations, the 
flux (and the fluence) are always integrated in the relevant  energy range.
Let us repeat that the estimations in the table are not supposed to be precise, they  should only convey the correct order of magnitude of the number of events. 

If we know the average distance of production of the flux, $D$, and when the emission is isotropic--to some degree of approximation--we can connect the observed  flux with the {\em intensity} of emission $I$ (i.e., number of neutrinos per second) namely,
\begin{equation}
\Phi=\frac{I}{4 \pi D^2}
\end{equation}
In this case, the total power radiated in neutrinos--in astrophysical parlance, the {\em luminosity}--will be
\begin{equation}
\mathcal{L}=I \times \langle E \rangle
\end{equation}
where we introduced the average energy of the neutrinos, $\langle E\rangle$.
If the emission is not isotropic, we have to replace $4\pi$ with 
$\Omega$, the solid angle of emission.
We proceed by illustrating with elementary considerations a couple of entries of table~\ref{tab1}.

\subsection{First example: pp-solar neutrinos}
The measured solar luminosity (of photons!) is 
\begin{equation}
\mathcal{L}_\odot =4\times 10^{33} \ \mbox{erg/s}
\end{equation}
this is in ultimate analysis due to the fusion of 4 hydrogen nuclei into helium, by a series of reactions that leads to
\begin{equation}
4 p\to \mbox{He}+2 e^++ 2 \nu_e
\end{equation}
and liberates $Q=26.7$ MeV, of which neutrinos take $\sim 1$\%. 
The 
number of reactions per second is
\begin{equation}
\dot{N}\equiv    \mathcal{L}_\odot /Q= 10^{38} \mbox{/s}=I/2
\end{equation}
where the last equation says that we have twice as many electron neutrinos. Distributing them over a sphere, we estimate  
a flux at Earth $\Phi\approx 6\times 10^{10}$ electron neutrinos per cm$^2$ per second. 

The reaction used in Borexino for their detection is the {\em elastic scattering} 
\begin{equation}
\nu+e \to \nu+e
\end{equation}
Its cross section can be estimated as 
\begin{equation}
\sigma_{\mbox{\tiny ES}}\sim G_F^2 E^2_\nu \sim 10^{-45}\ \mbox{cm}^2
\end{equation}
It is a direct prediction of the standard model and it depends upon the flavor of the impinging neutrino; see e.g.\ \cite{review}.
Note that when the neutrino energy $E_\nu$ is much larger than
the electron mass $m_e=0.5$ MeV, the behavior of $\sigma_{\mbox{\tiny ES}}$ with the energy changes from $E^2_\nu\to m_e E_\nu$.

\begin{figure}[t]
 \begin{minipage}[c]{9cm}
 \caption{\small\em By counting the number of neutrons produced in the dissociation of deuterium, due to  neutrino-induced
   neutral-current reactions, we can measure the flux of solar  neutrinos of any type. The SNO collaboration performed successfully this type of experiment and found a result in agreement with the theoretical predictions.}
   \label{deup}
   \end{minipage}
   \hskip5mm
  \begin{minipage}{8cm}
   \includegraphics[width=.73\linewidth]{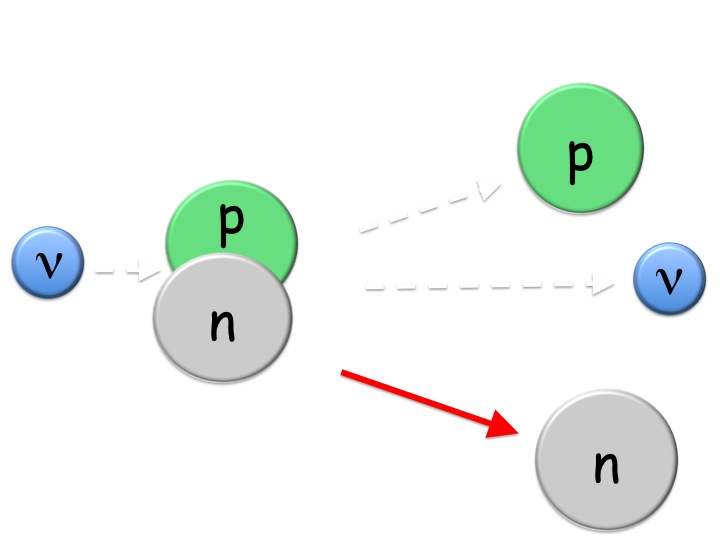} 
   \end{minipage}
\end{figure}
  
\subsection{Second example: supernova neutrinos}
 In order to allow the formation of a neutron star, the energy that should be radiated is 
 \begin{equation}
\mathcal{E}= G_N \frac{M_{\mbox{\tiny ns}}^2}{R_{\mbox{\tiny ns}}}\sim \mbox{const.}\times 10^{53}\mbox{ erg} 
 \end{equation}
 where we have the Newton constant $G_N=7\times 10^{-8}$ erg cm/g$^2$, the mass of the star
 $M_{\mbox{\tiny ns}}=1.4 {M_\odot}=3\times 10^{33}$ g, the radius of the star $R_{\mbox{\tiny ns}}=15 $ km; then, the number `const.' in previous estimation is 4, a somewhat more  accurate value used in the following is 3. 
 This energy is carried away by the six types of neutrinos; 
 if they have an average energy of 12 MeV, this will mean about few $10^{57}$ electron antineutrinos. Thus, we can equate it to the average power over the time of emission: 
 \begin{equation}\mathcal{E}=6\, T\, \mathcal{L}\end{equation} 
 where $\mathcal{L}$ is the power emitted per type of neutrino,
 $T\sim 10$ s is the time of emission, and 6 
 are the types of neutrinos and antineutrinos. 
 If they are emitted at a typical Galactic distance of 10 kpc, the expected fluence per type of neutrino is 
 \begin{equation}
 F=2 \times 10^{11}\ \mbox{cm$^{-2}$}
 \end{equation}
 The cross section of detection is the {\em inverse beta decay}, entailing electron antineutrino interactions on hydrogen nuclei,
  \begin{equation}
 \bar\nu_e + p \to e^+ + n
 \end{equation}
  A precise expression of the cross section is in \cite{ibdx}; 
 it can estimated as 
 \begin{equation}
 \sigma_{\mbox{\tiny IBD}} \sim G_F^2\ p_e E_e 
 \mbox{ with }E_{e^+}=E_\nu- (m_n-m_p)
 \end{equation}
In pure scintillators such as Borexino, the positron and the neutron are both observable. 
 The energy released 
 from positron annihilation, $2 m_e$, adds to the 
 kinetic energy of the positron 
 $E_{e^+}-m_e$.

{\small
\vskip4mm
\leftline{\bf Exercises of Sect.~\ref{s2}}
\vskip1mm
\noindent \sf 
4) Compare $\sigma_{\mbox{\tiny ES}}$ and  $\sigma_{\mbox{\tiny IBD}}$ cross sections. Calculate them in the standard model.\newline
5) Choose one  of the (or more) lines of  table~\ref{tab1} and check the calculations in some detail.\\
6) Use the values of the solar luminosity and of the solar 
temperature, about 6000 K, to estimate the solar radius by thermodynamics. Repeat the same steps to predict the inner radius of the supernova from where neutrinos but not light quanta can escape freely
({\em neutrino-sphere}).\\
7) Estimate the number of neutrinos and antineutrinos from a supernova and compare these numbers with the number of electrons originally present.

}

 \section{Neutrinos oscillations \label{s3}}
One of the first things one should know (and that, plausibly, most  readers know already) about neutrino observations is that, in many cases, there is a severe disagreement between the measured and the predicted neutrino fluxes. An important example is the one of solar neutrinos, that 
have been predicted by Bahcall in sixties; in fact, all observations,  beginning with those of the Homestake experiment, found a flux of electron neutrinos that is systematically smaller.\footnote{Recall that 
the leader of Homestake, Ray Davis Jr.\ and the one of Kamiokande, Masatoshi Koshiba, were awarded the Nobel prize in 2002 {\em for the detection of cosmic neutrinos,} not for the discovery of oscillations. Note also that the disagreement is today evident in  the Borexino experiment, able to  observe solar neutrinos of all possible energies.}
A related crucial observation was made in 2002. The SNO experiment \cite{snop} used neutral current reactions to see solar neutrinos,  thereby counting {\em all} types of neutrinos and not only the electronic ones: see Fig.~\ref{deup}. Since their observation agrees with the predictions--i.e.\ there is no shortage of events--one concludes that neutrinos do not disappear, but rather, they partially   change type  along their trip to the Earth.

\begin{figure}[t]
    \begin{minipage}[c]{6.7cm}
     \caption{\small\em 
     A ray of light propagating in a  crystal changes its polarization by $90^\circ$ if the phase velocities of the horizontal and vertical components are different and the size is chosen appropriately~\cite{wiki}. The description in terms of elementary particles is more dramatic: upon propagation, each photon changes its nature  (i.e.\ each state is transformed into its orthogonal state). Free neutrinos are subject to similar transformations acting on flavor space (Sect.~3.1). In suitable situations, 
 the propagation in matter    produces additional effects (Sect.~3.2).}
   \label{spar}
   \end{minipage}\hskip8mm
    \begin{minipage}[c]{7.5cm}
   \includegraphics[width=1\linewidth]{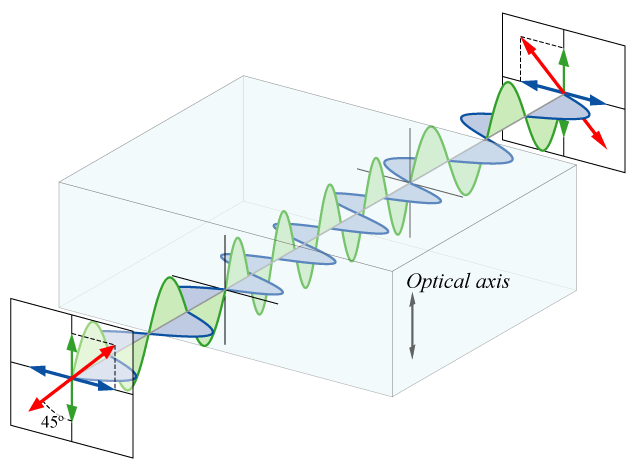} 
   \end{minipage}
\end{figure}

This can be explained by the occurrence of 
the phenomenon known as neutrino oscillation (aka flavor transformation) that is analogous to well-known phenomena, as the rotation of the polarization of the light in certain crystals, that is illustrated in Fig.~\ref{spar}. Below, we provide the reader with a basic description of this phenomenon; 
for a more complete account see~\cite{review}.

  \begin{figure}[t]
 \begin{minipage}[c]{4.8cm}
\caption{\small\em 
   Illustration of the mass spectra compatible with 
   the data from neutrino oscillations; left, normal hierarchy; right,
   inverted hierarchy. The minimum mass is not probed by oscillations.}
   \label{hie}
  \end{minipage}\hfil
    \begin{minipage}[c]{9cm}
   \includegraphics[width=1.1\linewidth]{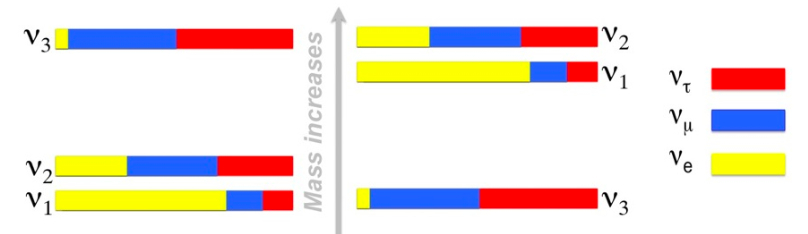} 
   \end{minipage}
\end{figure}

 \subsection{Basic formalism (Pontecorvo)\label{s3.1}}
 This first idea was proposed in 1957 by B.~Pontecorvo. He postulated the possibility that neutrinos could become antineutrinos, in analogy with what happens in the particle physics world.\footnote{In 1955 Gell-Mann and Pais undestood that weak interactions mix $K^0$ and $\bar{K}^0$, removing the mass degeneracy.   If one of them is produced at rest,  after some time $t$ 
 its two components with masses $m_1\neq m_2$ will acquire different phases $\exp(-i m_i c^2 t/\hbar)$ causing a partial transformation into the other 
 particle. This was observed in 1956.}
 Ten years later, before solar neutrino results were known, 
 Pontecorvo refined this proposal and suggested that this conjecture was valid for neutrinos of different species. The starting point is the formalism elaborated in 1962 by Maki, Nakagawa and Sakata  in Nagoya, that allows us to write the known neutrino states as a superposition of mass eigenstates. Let us consider the simple 
 case with two states,
 \begin{equation}
 \left\{
 \begin{array}{l}
 |\nu_e\rangle= +\cos\theta |\nu_1\rangle + \sin\theta |\nu_2\rangle \\
|\nu_\mu\rangle=- \sin\theta |\nu_1\rangle + \cos\theta |\nu_2\rangle 
  \end{array}
  \right.
 \end{equation}
 Suppose to produce a state $|\nu_e\rangle$, that,  
 in the moment of the production is orthogonal to the state  
 $|\nu_\mu\rangle$. Now, suppose to have 
 a distribution over the momentum (=a wavepacket)  
 and consider the  two mass components that have the same momentum. 
 Their propagation (=de Broglie's) phases,  
 \begin{equation}\exp\left[i\, \frac{\vec{p}\, \vec{x}-E_i\, t}{\hbar}\right]\mbox{ with } E_i=\sqrt{\vec{p}^2+m^2_i}\end{equation} 
 will become different  in the course of the propagation since 
 $E_1\neq E_2$. The difference of phase is,
  \begin{equation}
E_1-E_2=\frac{E_1^2-E_2^2}{E_1+E_2}=
\frac{m_1^2-m_2^2}{E_1+E_2} \approx
\frac{m_1^2-m_2^2}{2 E}
 \end{equation}
 where in the last step, we used the assumption of ultra-relativistic neutrinos, $E\approx |\vec{p}|\gg m_i$ and we set $c=1$ for convenience. 
 Thus, $ |\langle \nu_e,0 |\nu_e,t\rangle|\neq 1$ at some $t$, or in other words,  
 the electron neutrinos will not survive as such; conversely, there will be a finite probability that it will be turn into a muon neutrino; and vice versa. It is not difficult to derive the following formulae, 
\begin{equation}
 \left\{
 \begin{array}{lr} \label{pisco}
P_{\nu_e\to \nu_e}\equiv |\langle \nu_e,0 |\nu_e,t\rangle|^2 =
1-\sin^2 2\theta \sin^2\!\left[ \frac{m_2^2-m_1^2}{4 E} L\right] & \mbox{\small [disappearance/survival probability]} \\[2ex]
P_{\nu_e\to \nu_\mu}\equiv |\langle \nu_\mu,0 |\nu_e,t\rangle |^2=
\sin^2 2\theta \sin^2\!\left[ \frac{m_2^2-m_1^2}{4 E} L\right]
 & \mbox{\small [appearance probability]} 
  \end{array}
  \right.
 \end{equation}
 where $L$ is the distance of propagation $L\approx c t$ and the energy in this formulae is just the kinetic one $E\approx |\vec{p}|$ (and where we use $\hbar=c=1$). As a check, we note that summing the two probabilities, we find 1, that means that  neutrinos do not get lost; they just change type during the propagation. 
 Today, the leptonic mixing matrix (more on that later) is called PMNS mixing to honor the pioneering contributions of these scientists.\footnote{Note that: the two words  `virtual transmutation' in the English translation of the Nagoya paper suggest that they had some insight of the phenomenon we are discussing;
the hypothesis of `mixed neutrinos' is anticipated by a team in Tokyo, Katayama et al, 1962;
the circulation of ideas was much less efficient than today.} 
The story is beautifully recounted in the review paper \cite{bp}.

\begin{figure}[t]
   \centering
\includegraphics[width=.45\linewidth]{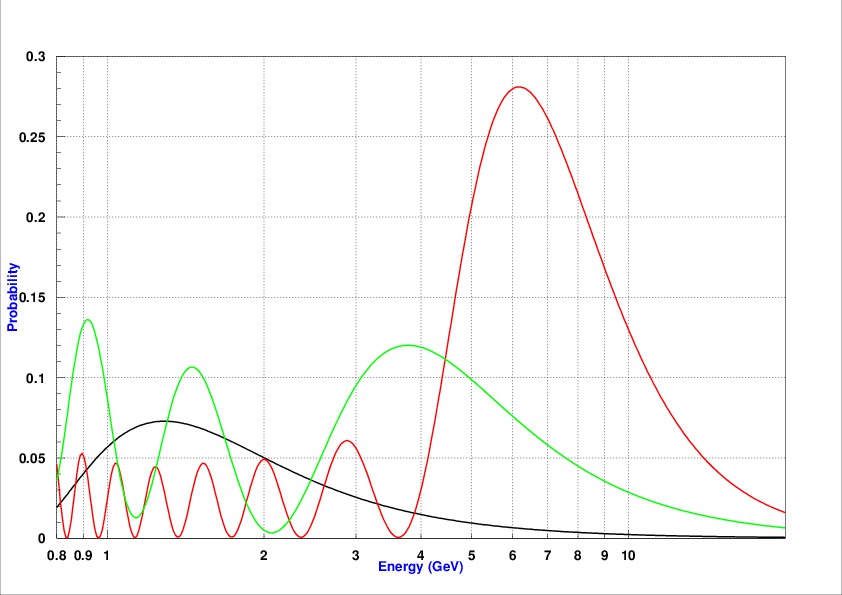}\hfill\includegraphics[width=.45\linewidth]{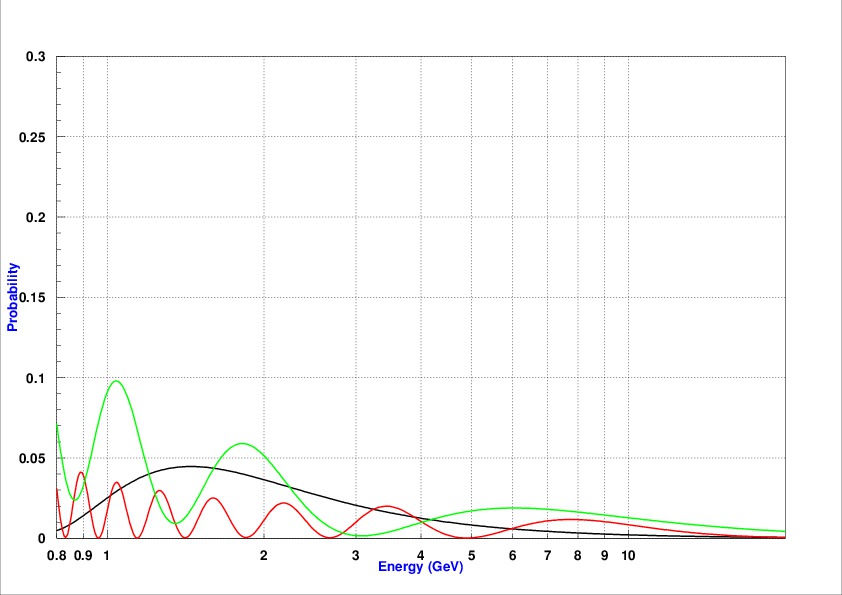} 
  \caption{\small\em 
   Matter effect  for normal mass hierarchy. 
Left panel:  
The probability of oscillation  $P_{\nu_\mu\to \nu_e}$
for a distance of 730 (black), 2500 (green), 6371 (red) km. Right panel: the same but for $P_{\bar\nu_\mu\to \bar\nu_e}$. The case of inverted hierarchy corresponds, in good approximation, simply to a  swap of the two panels.
  }
   \label{mss}
\end{figure}
 
 \subsection{Matter effect (Mikheev, Smirnov, Wolfenstein)\label{s3.2}}
 Vacuum oscillations occur only when the phases between the mass eigenstates depart from them. 
 But there are other phases that modify the transformation of neutrinos.  In fact,  the abundant presence of electrons in ordinary matter 
 provides the electron neutrinos with a peculiar phase of scattering. This corresponds to the forward scattering of neutrinos due to 
 weak interaction hamiltonian, 
 \begin{equation}
 H=
\frac{G_F}{\sqrt{2}} \int d^3 x\ \overline{\nu_e} \gamma^\mu (1-\gamma_5) \nu_e\ \langle \bar{e} \gamma_\mu (1-\gamma_5) e \rangle
 \end{equation}
 where the average is on the electron state. The only non-zero term for ordinary matter--i.e.\ unpolarized matter at rest--is $\langle \bar{e} \gamma_0 e\rangle=
 \langle {e}^\dagger e\rangle\equiv n_e$, namely, the density of electrons (measured e.g.\ in electrons per cm$^3$)  that equates to the molar density $\rho_e$ times the Avogadro number. The phase of scattering of $\nu_e$ in a sample of matter of size $dx$ 
 is then given by $\sqrt{2} G_F n_e dx$.   This term leads to matter effect on neutrino oscillations--in short, {\em matter effect} or {\em MSW} effect after Wolfenstein~\cite{msw1}, 
 Mikheev and Smirnov~\cite{msw2}.
 The size of the new phase can be  compared with the vacuum phase as follows,
 \begin{equation}
 \varepsilon\equiv 
 \frac{\sqrt{2} G_F n_e}{\Delta m^2/(2 E_\nu)}
 \approx \left\{
\begin{array}[c]{l}
 \left(\frac{7.5\times 10^{-5} \mbox{eV}^2}{\Delta m^2} \right)
 \left(\frac{E_\nu}{5\ \mbox{\footnotesize MeV}} \right)
  \left(\frac{\rho_e}{100\ \mbox{\footnotesize mol/cm}^3} \right)\\[2ex]
 \left(\frac{2.4\times 10^{-3} \mbox{eV}^2}{\Delta m^2} \right)
 \left(\frac{E_\nu}{5\ \mbox{\footnotesize GeV}} \right)
  \left(\frac{\rho_e}{3\ \mbox{\footnotesize mol/cm}^3} \right)
  \end{array}
  \right.
\end{equation}
where $\Delta m^2$ stands for a difference of squared masses.
E.g., let us focus on the solar neutrinos 
(for which $\Delta m^2=7.5\times 10^{-5}$ eV$^2$ and that are 
produced where $\rho_e\sim 100$ mol/cm$^3$).
When the energy $E_\nu\ll 1$ MeV, e.g.\ for the pp-neutrinos, the matter effect  can be neglected; instead,  for Boron neutrinos with energies 
above 5 MeV this is important. 
The effect exists  also for high energy neutrinos that cross the Earth
(when $\rho_e\sim 3$ mol/cm$^3$ and $\Delta m^2=2.4\times 10^{-3}$ eV$^2$) but this still to be observed.  In fact, 
it depends on the  arrangement of the neutrino spectrum, that is  unknown to date: The two possibilities are shown in Fig.~\ref{hie}.  
The relevant analytical formulae, obtained assuming constant matter density, 
give a reasonable approximation of the result, e.g.
\begin{equation}
P_{\nu_\mu\to \nu_e}=\sin^2\theta_{23}\left( \frac{\sin2{\theta_{13}}}{\Xi} \right)^{\! 2} \sin^2\left[ \frac{{\Delta m^2} L}{4 E} \Xi \right] \mbox{ with }
\Xi= \sqrt{  \sin^22\theta_{13}  +
(\cos2\theta_{13}- \varepsilon)^2 }
\end{equation}
Here we used the sign of $\varepsilon$ for the case of normal hierarchy; this has to be flipped for inverted hierarchy (left plot of 
Fig.~\ref{hie}).
Note that when $\epsilon=0$, we have
$\Xi=1$, and this formula remembers 
closely the vacuum formulae in 
Eqs.~\ref{pisco}.\footnote{Actually, it coincides with the vacuum formula
$|\ U_{e3} U_{\mu3}\times [\exp({\Delta m^2} L/{2 E}) -1 ]\ |^2$, written 
with the conventional choice of the mixing elements
$U_{e3}=\sin\theta_{13}$ and $ U_{\mu3}=\cos\theta_{13}\sin\theta_{23}$, see e.g.\ \cite{review}.}
The probabilities of oscillations can be calculated at the web site 
\cite{webo} that runs the code described in \cite{rubbia}; a sample output is in Fig.~\ref{mss}.
For more discussion on matter effect see also \cite{review} and the appendix of \cite{LujanPeschard:2013ud}.

 \begin{figure}[t]
   \centering
   \includegraphics[width=0.9\linewidth]{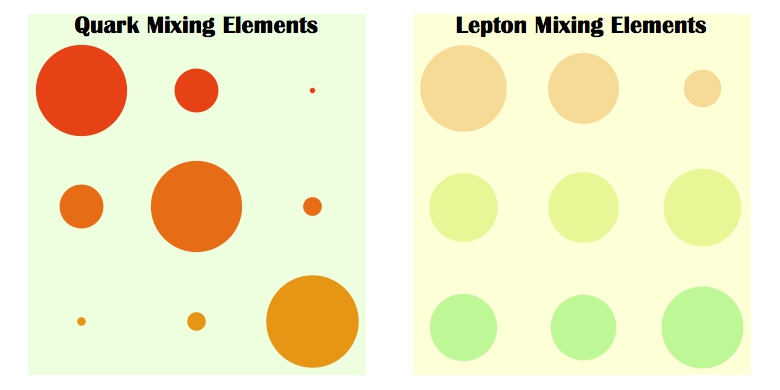} 
   \caption{\small\em 
   The surfaces of the circles represent  the size of the mixing elements. From top to bottom,  from left to right:
   Left panel, quark mixing (CKM) elements $|V_{ud}|,  V_{us}|,
   V_{ub}|,|V_{cd}|, ...$;
   Right panel, lepton mixing (PMNS) elements $|U_{e1}|,  U_{e2}|,
   U_{e3}|,|U_{\mu 1}|, ...$ 
      in both panels, the mixing matrices are supposed to be unitary.  
   The hierarchical structure of quark mixing elements contrasts with the one of lepton mixing elements.}
   \label{mixing}
\end{figure}

\subsection{What do we know on neutrino masses and mixings?}
Starting from the standard model, the most natural hypothesis is to assume that we have just three light neutrinos.\footnote{This is supported by the width of the $Z^0$, that measures the number of weakly interacting neutrinos, but also by  measurements based on the big-bang cosmology, namely:  the primordial nucleosynthesis of light elements and the distribution of the inhomogeneites of the cosmic microwave background.} We can take into account  the evidences of oscillations by postulating that they have mass and mix among them. The two types of mass arrangements (aka hierarchies or ordering or spectra) compatible with the present data  are illustrated in Fig.~\ref{hie}. 
This figure shows that each of the neutrino mass eigenstates $\nu_i$ is a superposition of the flavor states $\nu_\ell$. 
The connection is given by the leptonic (or PMNS) mixing matrix, that links these two kinds of field,
\begin{equation}
\nu_\ell=\sum_{i=1}^3 U_{\ell i}\ \nu_i\mbox{ with }\ell=e,\mu,\tau 
\end{equation}
A quantitative information is given in Fig.~\ref{mixing}, where we compare the sizes of the elements of the leptonic and of the quark mixing matrices, $|U_{\ell i}|$ and $|V_{ud}|$. Recall that the physical parameters are 3 mixing angles and 
one phase that describes CP violating phenomena. All 3 leptonic mixing angles are known in good approximation and we have the first hints that the CP violating phase is different from zero; but, let us repeat, the mass hierarchy is unknown to date.

Can we determine the mass hierarchy
by means of oscillations? 
Considering 3 flavor oscillations, the survival probability 
$P_{\bar\nu_e\to\bar\nu_e}$ of reactor antineutrinos is not the same for the two mass hierarchies shown in 
Fig~\ref{hie}. The effect is maximum at about 50 km, but it is not large and requires very precise measurements; this is the target of JUNO. The study of Earth matter effect via $P_{\nu_\mu\to \nu_\mu}$, $P_{\nu_\mu\to \nu_e}$, 
$P_{\nu_e\to \nu_\mu}$, $P_{\nu_e\to \nu_e}$
and connected antineutrino channels (see Fig.~\ref{mss})
offers other possibilities, that can be pursued already with atmospheric neutrinos (PINGU, ORCA, INO, Super- and Hyper-Kamiokande).
Also long-baseline experiments such as  NO$\nu$A, T2K (possibly with Hyper-Kamiokande) can study the mass hierarchy and the CP violating phase. 

\begin{figure}[t]
 \centering
\includegraphics[width=0.9\linewidth]{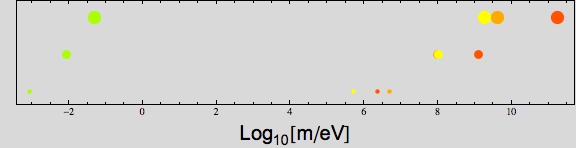} 
   \caption{\small\em 
  Masses of the known elementary spin 1/2 fermions.
   Red, up quarks; yellow, down quarks; orange,  
   charged leptons; green, neutrinos. In the last case, we assume normal mass hierarchy and 
   a mass ratio of 1/10 between the two lightest neutrinos
   for illustration purposes.}
   \label{masse}
\end{figure}

Finally, we would like to discuss neutrino masses.  
A quantitative illustration of the spectrum, assuming normal mass hierarchy, is in Fig.~\ref{masse}.
Note that oscillations fix only the squared mass differences without  providing us information on lightest neutrino mass. Other experiments can give us information on some  combination of neutrino mass. These include cosmological measurements of the distribution of the matter,  the search 
of mass effects in the (endpoint of) beta spectrum, the search for the lepton number violating transition  named neutrinoless double beta decay that receives a contribution from Majorana neutrino masses (see Sect.~\ref{impli}).  
The present 95\% CL bound on the sum of masses from cosmology is  140 meV \cite{viel}.
This bound  
allows us to deduce other stringent bounds,
\begin{quote}
\begin{tabular}{c||c|c|c} 
 & \bf theory & \bf beta spectrum  & \bf  neutrinoless $\mathbf{\beta\beta}$\\ 
parameter & $m_{\mbox{\tiny lightest}}$ & $m_{\beta}$ & $m_{\beta\beta}(\mathrm{\tiny max})$ \\ \hline\hline &&& \\[-2ex]
\sc normal hierarchy & 38 meV & 39 meV & 39 meV  \\[1ex]
\sc inverted hierarchy & 28 meV & 56 meV & 55 meV
\end{tabular}
\end{quote}
\noindent All these quantities are below the sensitivity of near future experiments. If the bound is correct, a positive detection in laboratory experiments ($\beta$ or $\beta\beta$) would point to physics beyond the minimal scenario; e.g.\  new sources of lepton number violation beyond light neutrino masses  contributing to 
neutrinoless double beta decay. 
Note that if we treat the cosmological bound as Gaussian, we find at 1 sigma that the sum of mass is below 71 meV; thus, the normal mass hierarchy is slightly favored by the interpretation of 
cosmological observations. 

%
%

{\small
\vskip2mm
\leftline{\bf Exercises of Sect.~\ref{s3}}
\vskip1mm
\noindent \sf 
8) Derive the formula of the survival and of the appearance probabilities; note that in order to obtain it, a crucial hypothesis is to consider the case of {\em ultra-relativistic neutrinos}. [This is the only case that applies in practice, though it could be interesting for theoretical--or academic--purposes to consider the other one.]\\
9) Including $\hbar$ and $c$ factors in Eqs.~\ref{pisco}, prove the important numerical formula
\begin{equation}
\varphi\equiv 
\frac{\Delta m^2 L}{4 E}=1.267 \frac{\Delta m^2}{\mbox{eV}^2}
\frac{L}{\mbox{km}}\frac{\mbox{GeV}}{E}
\end{equation}
that allows to estimate easily 
the cases when oscillations are absent ($\varphi\ll 1$) when 
they exist in proper sense ($\varphi\sim 1$) and when they  
are described by a constant factor instead ($\varphi\gg 1$).\\
10) Check that the maximum of the red curve  
in the left panel of Fig.~\ref{mss}, due to 
matter effect, corresponds to the region where the phase of scattering due to matter is comparable to the phase due to vacuum oscillations.\\ 
11) Knowing that the mass measured in cosmology is 
$\Sigma=m_1+m_2+m_3$ and using the values of the mixing elements and of the differences of mass squared, derive the bounds on  the masses  defined as $m_\beta^2=\sum_i |U_{ei}^2| m_i^2$
and $m_{\beta\beta}=|\sum_i U_{ei}^2 m_i|$ (the last one implying 
$m_{\beta\beta}(\mathrm{\tiny max})=\sum_i |U_{ei}^2| m_i$).\footnote{
Applying the Cauchy-Schwarz 
inequality
$|\langle \vec{x},\vec{y}\rangle |
\le |\!| \vec{x} |\!|\  |\!| \vec{y} |\!|$
to the vectors with components 
$(\vec{x})_i=U_{ei}^*$ and $(\vec{y})_i=U_{ei} m_i$ 
we have $m_{\beta\beta}\le 
m_{\beta}$, since $|\!| \vec{x} |\!|=1$.
Note 
that $m_{\mbox{\tiny lightest}}=m_1$ (=$m_3$) in normal (inverted) hierarchy, see  Fig.~\ref{hie}.}}

  \section{Modifying the standard model to include neutrinos masses
  \label{s4}}
  In this last section, we discuss a modification of the standard model of elementary particles, that allows us to take into account neutrinos masses. In this manner, we can explain oscillations and discuss on firmer bases  new   phenomena.
  \subsection{The fermions of the standard model}
\begin{figure}[t]
 \begin{minipage}[c]{7.5cm}
\caption{\small\em 
   In the standard model, each of the three families contain 15
   spin-half chiral (Weyl) states, including one neutrino with left chirality. 
   The corresponding  antiparticles  are automatically included by the general principle of relativistic quantum field theory.
   In the figure, we show the states of the first family, that  
   corresponds to the following mass eigenstates: $u$ and $d$ quarks (charged and colored), the lepton $e$ (charged)
   and neutrino $\nu_e$ (neutral and massless).}
   \label{fu}
  \end{minipage}\hskip10mm
    \begin{minipage}[c]{6.5cm}
   \includegraphics[width=1\linewidth]{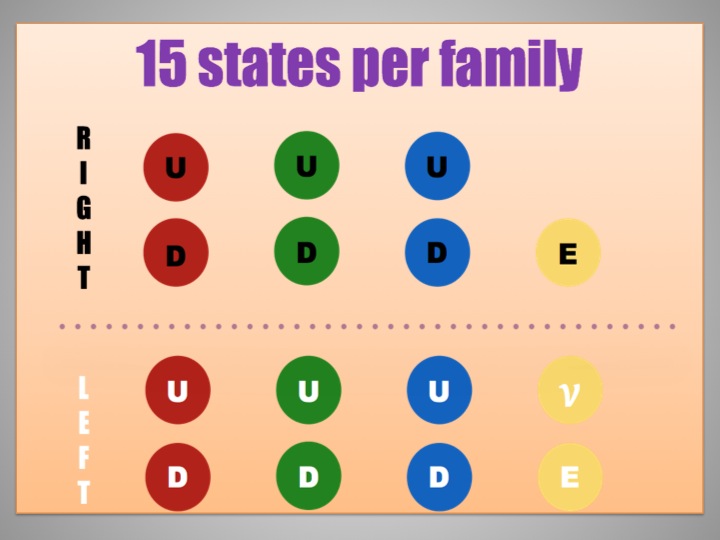} 
   \end{minipage}
\end{figure}

 Let us examine the meaning of neutrino masses from the point of view of the
 standard model. The latter is based on the gauge group 
  \begin{equation}
  \mathcal{G}_{\mbox{\tiny SM}}=
  SU(3)_{\mbox{\tiny color}} \times 
  SU(2)_{\mbox{\tiny left}}\times 
  U(1)_{\mbox{\tiny hypercharge}}
  \end{equation}
  and it includes 3 families of fermions, each one with 2 quarks (coming in 3 colors)  and 2 leptons, see Fig.~\ref{fu}.  It allows the transitions between the 3 different  families of quarks but it forbids those between different families of leptons. However, the observation  of neutrino oscillations/flavor conversion shows that this possibility does occur, just as it occurs for quarks. The most reasonable explanation, suggested in the previous pages, is that neutrinos have mass and mix among them. Thus, we ask the question of how to introduce neutrino masses in a suitable extension of the standard model.

The minimal modification (by definition) is to require that we do not add any new light particle. In this case, if we want to build  
some sort of neutrino mass, we have no other choice but 
to deal with the (left) leptonic doublet
  \begin{equation}
  L_{\ell a}=\left( \begin{array}{c} \nu_{\ell a} \\ \ell_a \end{array} \right)
  \mbox{ where }\ell=e,\mu,\tau\mbox{ and }a=1,2,3,4
  \end{equation}
since neutrinos are contained in this doublet of $SU(2)_{\mbox{\tiny left}}$.  The 
standard model quantum numbers of this doublet are 
$(1,2,-1/2)$ meaning that: it is a singlet under the color
group $SU(3)_{\mbox{\tiny color}}$ (of course! it is a lepton); it is a doublet under $SU(2)_{\mbox{\tiny left}}$ (even more evident) and 
it has  $Y=\rm{diag}(-1/2,-1/2)$,  and thus the electric charges are  $Q=\sigma_3/2+Y=\rm{diag}(0,-1)$, as it should be.
The Higgs doublet instead has the quantum numbers $(1,2,+1/2)$, thus it can be written $H=(H^+,H^0)$.

  \subsection{One additional operator in the standard model (Weinberg)}

We can form a gauge singlet by taking the product, $H i\sigma_2 L_{\ell a}=-\nu_{\ell a} H^0+\ell_a H^+$ 
but this bilinear combination is not Lorentz invariant, since it has one free spinorial index $a$. When the Higgs field takes vacuum expectation value, we get a term proportional to the neutrino field, 
 $ H i\sigma_2 L_{\ell a}=-\langle H^0\rangle \nu_{\ell a}+...$ It is sufficient to contract two of these terms to obtain an invariant term, that can be added to the standard model hamiltonian density, namely
 \cite{weinberg}
 \begin{equation} H \sigma_2 L_{\ell a}\ C_{ab}^{-1}\  H \sigma_2 L_{\ell' b} /(2 M) \label{wmb}
 \end{equation}
 where repeated indices are contracted. 
 Here, $M$ is a constant with dimensions of mass, that is included to ensure  that the hamiltonian density has the correct dimensions, while $C$ is the charge conjugation matrix, needed to form an invariant quantity out of two spinors. When the Higgs field takes vacuum expectation value we get a bilinear term built with the (left handed) neutrino fields only. This is, 
\begin{equation}
  \frac{m}{2}\  \nu_\ell C^{-1} \nu_{\ell'} 
  + \mbox{ hermitian conjugate, 
where }
m=\frac{ \langle H^0\rangle^2}{M}
\end{equation}
Note that the mass parameter $m$ is inversely proportional to the mass scale of the Weinberg operator, $M$.
We can generalize this position to include similar terms 
for all types of neutrinos by  replacing 
$m\to m_{\ell\ell'}$ where  $\ell,\ell'=e,\mu,\tau$. Finally, we define the following real (or Majorana) spinor:
\begin{equation}
\chi_\ell=\nu_\ell + C \overline{\nu_\ell}^t \mbox{ such that } 
\chi_\ell=C \overline{\chi_\ell}^t
\end{equation}
If $m_{\ell\ell'}$ is real, 
we can rewrite the above bilinear term in a manner that looks  familiar, namely
\begin{equation}\label{identicale}
\frac{m_{\ell\ell'}}{2}\  \nu_\ell C^{-1} \nu_{\ell'}+\mbox{hermitian conjugate}=
-\frac{m_{\ell\ell'}}{2}\overline{\chi}_\ell \chi_{\ell'}
\end{equation}
We conclude that the new operator produces a mass term for the neutrinos, called {\em Majorana mass}.

\subsection{Implications\label{impli}}

Evidently, the above possibility is particularly interesting, since (apart from the technicalities) it is compatible with the standard model gauge symmetries. Moreover, it gives rise to lepton number violating phenomena, such as the {\em neutrinoless double beta decay}--namely the nuclear transition 
\begin{equation}(A,Z)\to (A,Z+2)+2 e^-\end{equation} 
that is forbidden in the standard model. Thus there is some connection between neutrino oscillations and other 
phenomena--simply because they both result from neutrino masses.

However, `there is no such thing as a free lunch'. What is the price that  we have to pay if we  correct for the shortcomings of the standard model by adding the new operator of Eq.~\ref{wmb}? The fact is that it has canonical
(mass) dimension 5, thus it 
breaks the renormalizability of the resulting quantum field theory. 
This is not so dramatic as it might look, though; the same happens with Fermi interactions. In fact, $G_F$ has dimensions of inverse mass squared  (recall Eq.~\ref{fdef}) 
but this is far from meaning that Fermi interactions have no practical applications in nuclear and particle physics! 

To summarize, we can hypothesize that neutrino masses are due to
 the  operator shown in  Eq.~\ref{wmb}, but we 
are not yet ready to discuss its {\em origin};
we have to postpone the question of how it emerges from a renormalizable theory (that extends the standard model) 
to a future and more complete model of the world of  elementary particles, that will be hopefully based on new data and information.
Unfortunately, it is not clear whether the scale of new physics will allow direct access by terrestrial accelerators--see the last exercise.

{\small
\vskip2mm
\leftline{\bf Exercises of Sect.~\ref{s4}}
\vskip1mm
\noindent \sf 
12) Assuming that the spinor $\nu$ has left chirality, $\nu=P_L \nu$,
check that the spinor $C \overline{\nu}^t$ has right chirality.\\
13) Prove that the spinor $\chi$ has to be a trivial representation of any 
$U(1)$ group, and in particular, it  cannot transform under the 
$U(1)_{\mbox{\tiny e.m.}}$ of the electromagnetism--it cannot 
be a charged field. What about 
$U(1)_{\mbox{\tiny hypercharge}}$?\\
14)  Show that the two mass terms written in Eq.~\ref{identicale} are identical.
Prove that, in general, the matrix $m_{\ell\ell'}$ is symmetric.
Write the mass term using the Majorana field $\chi_\ell$, assuming  that 
 $m_{\ell\ell'}=\mbox{Re}(m_{\ell\ell'})+i\ 
 \mbox{Im}(m_{\ell\ell'})$ is complex.\\
15) Show by direct calculation 
that the combination $H \sigma_2 L$ is a gauge singlet of the standard model gauge group, and in particular it is 
invariant under $SU(2)_{\mbox{\tiny left}}$ transformations.
What are the transformation properties of $H \sigma_2 \vec{\sigma} L$?
Combine the $SU(2)_{\mbox{\tiny left}}$ spinors $L$ and $H$
in other manners 
and show that the result is always proportional to the  
Weinberg operator. (Recall that $2\otimes 2=1\oplus  3$.)\\
16) Discuss which are the values of masses $M$ in Eq.~\ref{wmb} 
 that are compatible with the masses of neutrinos observed by mean of oscillations.  Discuss the meaning on $M$ by elaborating on the correspondence between Eqs.~\ref{fdef} and
 ~\ref{wmb} possibly by envisaging a specific renormalizable model to account for  this operator.

}

 
 \acknowledgments{I would like to thank A.~Ianni and the Organizers for the invitation, and P.~Sapienza for careful reading of the text. 
An appeal to potential readers: 
please feel free to contact me at to discuss this note, the exercises, or to get more information but 
 do not worry too much if your results do not agree with those in table~\protect{\ref{tab1}}; 
 they are only meant to allow first orientation.
 
 }
    

\ctable[
caption={\label{tab1}\small 
Abbreviations: (1) BX=Borexino; SK=Super-Kamiokande (22.5 kton);
 Ice$^3$=IceCUBE. 
 (2) ES=elastic scattering; IBD=inverse beta decay; QEL=quasi elastic nucleon interaction; DIS=deep inelastic scattering.
 (3) e=electron; p=proton; N=nucleon. 1 pc=$3 \times 10^{18}$ cm.
Question marks denote purely theoretical predictions.},
label={tab1},
botcap, 
sideways 
]
{c|ccccccc|c}
{
}
{
Neutrino & Flux & Average  & Target \# & Time & Distance &  Cross  & Remarks \& & Event \\ 
source  & [1/cm$^{2}$s] & energy  &  \& type  &  & [cm]  &  section [cm$^2$] & reference & \# \\ \hline
Sun (pp) & $3\times 10^{10}$ & 0.2 MeV &  $3\times 10^{31}$ e & 1 day & $1.5\times 10^8$  km 
  & $10^{-45}$ {\small (ES)} & 
BX \cite{pp}
& 100 \\
Earth & $3\times 10^{5}$ & 2 MeV &  $ 10^{31}$ p & 4 yr & ? & $3\times 10^{-44}$ {\small (IBD)} & BX \cite{geon}
& 10 \\
Reactor & $3\times 10^{4}$ & 4 MeV &  $ 10^{31}$ p & 4 yr & 1000 km 
 & $ 10^{-42}$  {\small (IBD)} & BX \cite{geon}
& 30 \\
Relic SN & $2.5 $ & 10 MeV &  $ 10^{33}$ p & 1 yr & 300 kpc 
& $7\times 10^{-42}$  {\small (IBD)} & SK  \cite{relic}
& 0.6? \\
Galactic SN & $10^{10} $ & 20 MeV &  $ 10^{33}$ p & 10 s & 10 kpc 
& $3\times 10^{-41}$  {\small (IBD)} & SK \cite{sn}
& 5000 \\
Atmosphere & $1 $ & 1 GeV &  $ 10^{34}$ N & 1 yr & 500 km 
& $ 10^{-38}$   {\small (QEL+DIS)} & SK (all $\nu$)
& 3000 \\
Accelerators $\nu_\mu$ & $0.1  $ & 10 GeV &  $ 6\times 10^{32}$ N 
& 1 yr & 500 km 
& $ 10^{-37}$   {\small (DIS)} & SK 
& 3000 \\
Accelerators  $\nu_\tau$ & $0.01  $ & 120 GeV &  $ 10^{34}$ N 
& 3 yr & 700 km 
& $ 10^{-38}$   {\small (DIS)} & Opera 
& 6 \\
Galactic source & $ 10^{-12} $ & 3 TeV &  $2\times 10^{39}$ N & 1 yr & 1 kpc 
& $ 10^{-35} $   {\small (DIS)} & Ice$^3$ 
& 1? \\
HE neutrinos & $2\times 10^{-12} $ & 100 TeV &  $6\times 10^{38}$ N & 3 yr & ? & $ 10^{-34}$  
 {\small (DIS)} & Ice$^3$ \cite{nature}
& 10 \\
}

\tableofcontents

\end{document}